\documentstyle[12pt,epsf]{article}
\def\beq{\begin{equation}}   \def\eeq{\end{equation}}
\def\lsim{\mathrel{\rlap{\lower3pt\hbox{\hskip0pt$\sim$}}
    \raise1pt\hbox{$<$}}}         
\def\gsim{\mathrel{\rlap{\lower4pt\hbox{\hskip1pt$\sim$}}
    \raise1pt\hbox{$>$}}}         

\begin{document}
\begin{titlepage}

\begin{flushright}
TPI-MINN-99/03-T\\
UMN-TH-1739/99\\
hep-ph/9904021\\
\end{flushright}

\begin{center}
\baselineskip25pt

{\Large\bf 
 Tilting the Brane, or Some Cosmological Consequences of the Brane
Universe}

\end{center}

\vspace{0.1cm}

\begin{center}
\baselineskip12pt

{\large G. Dvali}

\vspace{0.1cm}
Physics Department, New York University, New York, NY 10003,

\vspace{0.2cm}

{\em and}

\vspace{0.2cm}
{\large  M.~Shifman} 

\vspace{0.1cm}
Theoretical Physics Institute, University of Minnesota, Minneapolis, 
MN 55455

\vspace{1cm}

{\large\bf Abstract} \vspace*{.2cm}
\end{center}

We discuss  theories in which the standard-model particles
are localized on a brane embedded in space-time with large
compact extra dimensions, whereas gravity propagates in the bulk.
In addition to the ground state corresponding to a straight infinite
brane, such theories admit a (one parameter) family of stable
configurations corresponding to branes wrapping with  certain
periodicity around the extra dimension(s) when one moves along a
noncompact coordinate (tilted walls). In the effective four-
dimensional 
field-theory picture, such walls are interpreted as one of the (stable) 
solutions with the constant gradient energy, discussed 
earlier~\cite{DS1,DS2}. In the cosmological context their energy  
``redshifts" by the Hubble expansion and dissipates slower then the 
one
in  matter or radiation. The tilted wall eventually starts
to dominate the
Universe. The upper bound on the energy density coincides with the 
present critical energy density. Thus, this mechanism  can become
significant any time in the future.
The solutions we discuss are characterized by a tiny spontaneous
breaking of both the Lorentz and rotational invariances.
Small calculable Lorentz noninvariant
terms in the standard model Lagrangian are induced. Thus, the tilted 
walls  provide a framework for the spontaneous
breaking of the Lorentz invariance.
\vspace{0.5cm}

\noindent
{\em To appear in L.B. Okun Festschrift, Eds.   V. Telegdi and K. 
Winter, 
to be
published by North Holland}

\end{titlepage}

\section{Introduction}

Nonobservation of  low-energy supersymmetry may be due to the
fact that we live on a non-BPS topological defect, or brane, embedded
in higher-dimensional space-time~\cite{DS1}. While the external 
space
may or may not be supersymmetric, the effective low-energy theory 
on 
the non-BPS brane is not supersymmetric. 

The idea of the brane Universe is especially motivated by the 
solution of 
the hierarchy problem through lowering of the fundamental scale of 
quantum gravity down to TeV~\cite{add,AADD} (see also 
\cite{lw,ddg}).
In the model suggested in Ref.~\cite{add} the standard-model 
particles 
are localized on a topological defect, or $3$-brane, embedded in  
space 
with large extra compact dimensions
of size $R$, in which gravity can propagate freely. In other words, 
the 
original $(4 + N)$-dimensional space-time is assumed to be split into
${\cal M}_4\times {\cal M}^\prime_N$, where ${\cal M}_4$ is the
 four-dimensional Minkowskian space while
${\cal M}^\prime_N$ is a compact manifold. We will refer to 
it as to the external space. 
Within the scenario~\cite{add} one can lower the fundamental scale 
of 
gravity $M_{{\rm P}_{\rm f}}$ down to  TeV, or so. The observed
weakness of gravity at  large distances is then due to a large volume
of the extra space $R\gg M_{{\rm P}_{\rm f}}^{-1}$. The relation 
between the fundamental ($M_{{\rm P}_{\rm f}}$) and four-
dimensional 
($M_{\rm P}$)  Planck scales  can be derived~\cite{add} by virtue of 
the  
Gauss law, 
\begin{equation}
M_{\rm P}^2 = M_{{\rm P}_{\rm f}}^{N + 2}\, V_N \, ,
\label{planckscale}
\end{equation}
where $V_N$ is the volume of the external space with radius $R$. 
Although this relation is valid for any values of $M_{{\rm P}_{\rm 
f}}$
and $R$, the case  most interesting phenomenologically is 
 $M_{{\rm P}_{\rm f}} \sim $ TeV.  (For $R\sim $ 1 mm, this implies 
that the
extra space is two-dimensional.) 

 The most important  ``technical" question to be addressed 
 is  dynamical localization
of the standard-model particles on the brane. In the field-theoretic
context the fermions can be localized due to an index theorem, as  
was
suggested in \cite{localization1/2}, whereas the localization of the
gauge fields requires  the outside medium  to be
confining \cite{localization1}. In particular, this implies that free
charges cannot exist in the bulk.

 On the other hand, in the string-theoretic context, the most natural
framework for the brane-world picture is through the $D$-brane 
construction
(for a review, see e.g. \cite{D-branes}). In this context the
standard-model
particles can be identified with the open string modes stuck on the 
brane,
whereas gravity comes from the closed string sector propagating in 
the
bulk \cite{AADD,ST,ignatios}. For the purposes of the
present paper the precise nature of  localization will be
unimportant, since we will exploit  the low-energy effective 
field-theory approach for which the high-energy nature of 
the brane is beyond ``resolution".

 In this paper we will present new theoretical observations regarding 
the
branes on the manifolds ${\cal M}_4\times {\cal M}^\prime_N$
where ${\cal M}^\prime_N$ is compact. Then we
discuss  some possible
cosmological consequences of the brane Universe with the low-scale 
quantum gravity.
We point out that these theories admit stable solutions which could 
manifest
themselves through a tiny spontaneous breaking of the  Lorentz and 
rotational
invariances.

In  four-dimensional field theory the  solutions above can be
interpreted as  a recently discovered new class of {\em stable} 
vacuum 
configurations
in supersymmetric theories, with the constant gradient energy, 
which
may or may not break
supersymmetry~\cite{DS1,DS2}. In the latter case these solutions
generalize the notion of the BPS saturation to infinite
values~\footnote{By the
infinite central charges we do not mean trivial infinities associated 
with
the area of the wall or the length of the string.}  of the
central charge~\cite{DS2}. Below we will show how these
solutions can naturally emerge in the brane Universe picture.

 Before proceeding, let us note that other possible
cosmological implications of branes can be due to ``brane inflation"
driven by a displaced set of branes \cite{braneinflation} or due to
nonconservation
of global quantum numbers in the brane Universe \cite{gigad}.
In a very different context
a class of  time-dependent cosmological solutions was
discussed \cite{ovrut} within the Ho\v{r}ava-Witten approach 
\cite{HW}. Some aspects of thermal cosmology were
considered in \cite{mmtr}. These issues are not directly related,
however, to the present work.

For definiteness assume  $(4 + N)$-dimensional
space-time to be
${\cal M}_4\times {\cal T}_N$, where ${\cal M}_4$ is 
 four-dimensional
Minkowskian space while
${\cal T}_N$ is an $N$-torus with the radius $R$. The standard-model
particles are the modes living on a $3$-brane embedded in this
 space-time.
Let the brane tension (the energy per unit three-volume) be $T$.
Throughout the paper we will assume that $T \sim M_{{\rm P}_{\rm 
f}}^4$ 
(the only fundamental mass scale in the theory). Obviously, $T > 0$ if 
the 
brane is to be regarded as a dynamical object (the wall), since 
otherwise 
it
will  be unstable. Then, in the absence of  matter and radiation, the
effective four-dimensional energy density $\rho_{\rm eff}$ has two
contributions: a positive one from the brane tension, and a (must-be)
negative one from the bulk
\begin{equation}
\rho_{\rm eff} =  T + \Lambda_{\rm bulk}\, V_N  \, ,
\label{lambda}
\end{equation}
as  seen in experiments at distances $\gg R$. 
Here $\Lambda_{\rm bulk}$ is the negative bulk cosmological 
constant.
In the lowest-energy state these two must conspire to cancel each 
other.
This is a usual fine-tuning of the cosmological constant, about which 
we
have nothing new to say. The lowest-energy state is achieved 
 in the above picture
 when the brane is straight, its ``surface" area is minimal per
unit three-volume of the space (Fig. 1). However, since the brane is a
dynamical object it can bend or curve in the external space and, in 
general,
it
will do so. It can be tilted with respect to ${\cal M}_4\times {\cal 
T}_N$.
Note that this curvature {\it a priori}
has nothing to do with  the
gravitational curvature of the whole (brane + bulk) effective
four-dimensional space in the Friedmann cosmological equations
emerging at distances $\gg R$. The latter
will be assumed to vanish. Bending  the brane in the external space, 
or
tilting, it will produce an energy excess, or an effective energy 
density of 
the  Universe,
which  can be estimated as
\begin{equation}
\rho_{\rm brane}  \sim M_{{\rm P}_{\rm f}}^4\, (R/r)^2\, ,
\qquad r\gg R\, ,
 \label{braneenergy}
\end{equation}
where $r$ is a typical curvature radius of the brane, {\em not} to be
confused with the Friedmannian curvature radius, which we take 
infinite. (In the case of the tilted brane $r$ is  its longitudinal
dimension, $r\sim L$.  In this case 
Eq.~(\ref{braneenergy})
can  be obtained as follows:
\beq
\rho_{\rm brane}  \sim M_{{\rm P}_{\rm f}}^4\, (\alpha)^2\, ,
\label{atwo}
\eeq
provided that the tilt angle $\alpha\sim R/L\ll 1$, see below.) This 
energy
provides an effective force resisting to bending; the force tends to  
straighten
out the brane. What would be the cosmological significance of this 
excess
energy?

\begin{figure}   
\epsfxsize=8cm
\centerline{\epsfbox{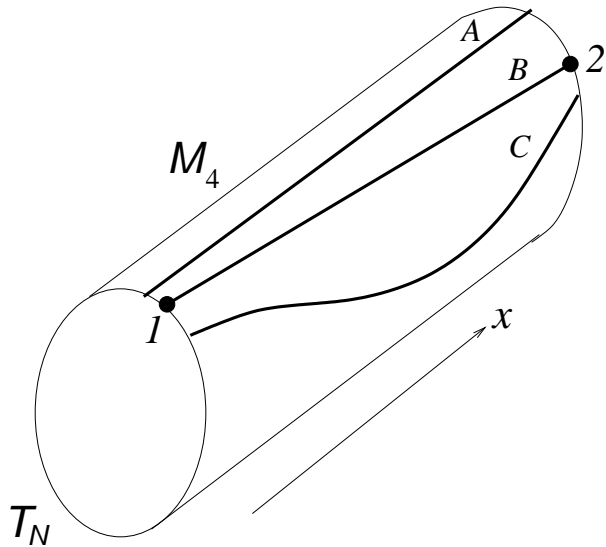}}
 \caption{Domain walls (branes) in ${\cal M}_4\times {\cal T}_N$.
$A$  -- the straight untilted brane, $B$ -- tilted brane,  $C$ -- curved 
brane. The length of the cylinder in the $x$ direction is $L\sim H^{-
1}$.}
\end{figure}

To estimate its impact  we have to know $r$. It is natural to assume 
that
in  the scale smaller than the Hubble size the 
brane is straightened out (no excessive crumpling). Whatever 
mechanism
solves the horizon and isotropy problems, it  would also help to this
straightening. So, it seems   reasonable to assume that
\beq
r^2 \gsim H^{-2} \gsim \rho_c^{-1}\, M_{\rm P}^2\, ,
\label{aone}
\eeq
where $\rho_c$ is the critical density of the Universe today.
This estimate obviously refers also to the  longitudinal
dimension  of the tilted brane, $L\gsim H^{-1}$. 
Now,  substituting this in Eq. (\ref{braneenergy})
and  using Eq. (\ref{planckscale}) with $N\geq 2$ we get 
\begin{equation}
\rho_{\rm brane} \lsim \rho_c  \, .
\label{energy2}
\end{equation}
The upper bound which can only, but not necessarily, appear for $N = 
2$
is intriguing since it suggests that the brane can serve as some sort of 
dark matter in  Universe. We will study the nature of its energy 
below.

The potential importance of the domain walls in the cosmological
considerations was recognized long ago~\cite{ZKO}.  In
Ref.~\cite{ZKO} it was shown
that  were the domain walls  within our universe, serious (potentially
terminal)  cosmological problems might arise in the theory. 
Situation  dramatically changes if our Universe itself is a wall.

\section{What is the tilted wall?}

To illustrate the idea we will consider, for simplicity,
the four-dimensional space  ${\cal M}_4$ compactified in ${\cal 
M}_3\times
{\cal T}$. The situation is general and does not depend on particular
details.

Let us start from ${\cal M}_4$ and assume that the dynamical theory 
under
consideration has  multiple discrete degenerate vacua. The simplest 
example
is the theory of the real scalar field with the potential of the double-
well 
type.
The simplest supersymmetric example is the 
minimal Wess-Zumino model with the
cubic superpotential. The field configuration depending only on one
coordinate (call it $z$) that interpolates between one vacuum at
$z\to-\infty$ and another at $z\to\infty$ is the domain wall. The 
width
of the wall $\delta$ in the $z$ direction is of order $\delta\sim M^{-
1}$
where $M$ is the mass scale of the field(s) of which the wall is made.
It is assumed that $\delta$ is much smaller than any other relevant 
scale of 
dimension of length.

On ${\cal M}_4$  it is meaningless to speak of the
tilted wall -- the direction $z$ can be chosen arbitrarily.
Now, we compactify $z$ and consider the theory on the
cylinder  ${\cal M}_3\times
{\cal T}$. The underlying dynamical theory must be
modified  accordingly,  in order to allow for the existence of the  
walls. 
In the
case at hand it is sufficient to assume that the wall-forming field 
$\Phi$ 
lives on a
circle, i.e. one can consider the model of the sine-Gordon type or its
supergeneralizations. Both, the superpotential and the K\"{a}hler
potential must be periodic in $\Phi$, with commensurate periods.
For simplicity we assume these periods to be $2\pi$.
Following an old tradition, we  rename the compact coordinate,
$z\to X$. The non-compact coordinates (including time) 
will be denoted by $x_\mu$. We look for the topologically nontrivial 
solutions
of the soliton type on the cylinder, depending on one coordinate only.
The solution of the type $\Phi_0 (X)$, which is independent of
$x_\mu$, is the straight wall, see $A$ in Fig. 1. 
It satisfies the condition $\Phi_0 (X+2\pi R)=\Phi_0 (X) +2\pi$. This 
wall 
 is aligned ``parallel" to the cylinder. The tilted wall  ($B$ in Fig. 1)
is a solution of the type $\Phi_\alpha ( X\cos\alpha +  x\sin\alpha)$ 
where
$\alpha$ is the tilt angle. Note that the function
$\Phi_\alpha$ does not coincide with $\Phi_0$, generally speaking.
 In the limit
of small $\alpha$ the difference between $\Phi_0$ and 
$\Phi_\alpha$ is
$O(\alpha^2)$.
 The condition  $\Phi_\alpha  ( (X+2\pi R)\cos\alpha +  x\sin\alpha)=
\Phi_\alpha  ( X\cos\alpha + x\sin\alpha )+2\pi$ must  be satisfied. 
It is 
not
difficult to see that the tilted wall solution exists, if so does the 
straight 
wall
solution. The  tilted wall is stable provided the solution is ``nailed" at 
the 
points
$1\,,\,\, 2$ at the boundaries (see Fig. 1). Alternatively, one can glue 
the
boundaries of the cylinder converting it into a two-torus. The tilted 
wall 
solution
is automatically stable then if the wall  winds around  the two-torus.

It is not difficult to prove that the domain walls on the cylinder 
cannot 
be
BPS-saturated, strictly speaking. However, if $\delta/R\ll 1$ the 
straight walls may be very close to the BPS saturation, achieving the 
BPS
saturation in the limit  $\delta/R\to 0$, when the wall becomes the
``genuine brane". The tilted walls are never
 BPS-saturated, their tension being larger than that of the straight 
wall.
This effect -- the increase of the internal tension $T$ of the
tilted wall compared to that of the straight wall -- is proportional 
both 
to
$\alpha$ and  to
$(\delta / R)^2$. It can be made arbitrarily small in the limit
$\delta / R \to 0$. We will neglect it in what follows, using one and 
the
same tension $T$ for the tilted and straight walls.  The tilting does 
produce an
impact on $\rho$ since the area of the tilted wall per unit length of 
the
cylinder in the $x$ direction  is larger.  This effect is most 
conveniently
described by the effective low-energy theory of the zero modes on 
the 
wall.
The corresponding discussion is presented in the next section.

\section{Calculating $\rho_{\rm brane}$ for non-vanishing tilt 
angles}

To deal with the long wave-length ($\lambda\gg \delta$) 
deformations
of the straight wall we will use
an effective four-dimensional field theory 
emerging for the zero modes. The wall solution spontaneously break 
the
translational invariance in one direction.  Correspondingly, in the 
simplest case
there arises one zero mode which is  the Goldstone boson of the 
spontaneously
broken symmetry. In more complicated models (see below) there 
may 
arise
several zero modes $\phi_A$. In the absence of gravity the effective 
Lagrangian
is
\beq
{\cal L} = M_{P_{\rm f}}^2 \,\int d^4x \, \frac{\partial 
\phi_A}{\partial 
x_\mu} \, 
\frac{\partial \phi_B}{\partial x^\mu} \, \eta^{AB}(\phi )\, ,
\label{dopone}
\eeq
where $\eta^{AB}$ is the external metrics depending on
the structure of the manifold on which the fields $\phi$ live. For 
instance, in 
the case of the domain wall in  Minkowski space
$\eta^{AB}=\delta^{AB}$. All dimensional  constants in the low-
energy 
theory of
the zero modes are related to the order parameter, which  
 is the brane tension  $T\sim M_{{\rm P}_{\rm f}}^4$. 
The dynamics of the Goldstone bosons  on the brane
is described~\cite{DS1} by $(3+1)$-dimensional field theory which 
may 
or may not be supersymmetric (in the latter case the brane must be 
BPS
saturated). If we deal with the supersymmetric theory  $\eta^{AB}$ 
is 
obtained
from the K\"ahler potential.

Let us assume for the time being that 
we have only one Goldstone boson and $\eta^{AB} =1$.
Then, the solution $\phi = \alpha x$ goes through the equations of 
motion of the
theory (\ref{dopone}). This is the constant energy density vacuum,
 discussed in
Refs. \cite{DS1,DS2}. To make contact with the discussion above, we 
note
that the vacuum $\phi =\alpha x$ represents the tilted wall
described by the solution $\Phi_\alpha$ in the full theory. The 
additional
contribution to $\rho$ compared to the straight brane  is obviously
\beq
\Delta\rho_{\rm brane} = \frac{T\alpha^2}{2}\, .
\eeq
This result has a very transparent interpretation in the full theory.
 It exactly
reproduces  the increase of the brane surface  per  unit 
length of the cylinder for a non-vanishing tilt
angle $\alpha$, see Fig. 2.  Given this interpretation, one might ask 
why
one needs to consider the effective low-energy theory at all.
The point is that at the next step we want to switch on gravity in the 
bulk. 
Having the low-energy theory of the zero modes, describing
matter on the brane, helps analyze the impact of gravity.

\begin{figure}   
\epsfxsize=8cm
\epsfysize=7cm
\centerline{\epsfbox{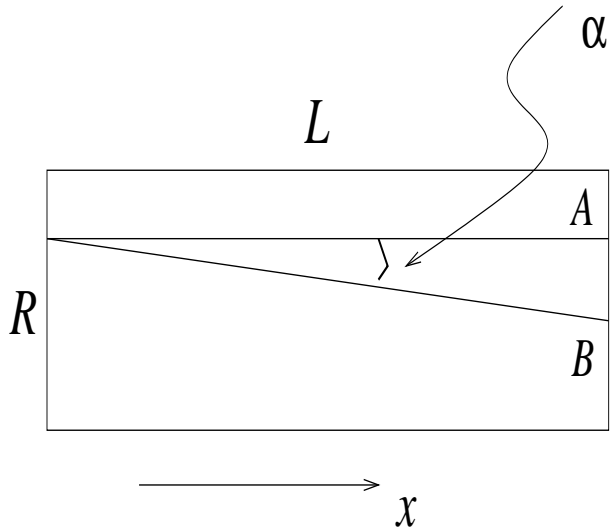}}
 \caption{The map of the cylinder of Fig. 1. The ratio of the surfaces 
of 
the
branes $B$ to $A$ per unit length in the $x$ direction is 
$1+\alpha^2/2$.}
\end{figure}

 In the language of the effective field theory~(\ref{dopone}), the 
bending 
 of the brane
is equivalent to $\phi^A$ acquiring some $x_{\mu}$ dependence in 
the
vacuum. The generic $\phi^A(x)$ configuration is unstable and will
decay producing
$\phi$ waves (the sound waves on the brane). 
This is the mechanism of eliminating foldings on the brane.
Since these waves  travel
with the speed of light, the brane will iron itself out in the
horizon scale. Eventually it will evolve to  a tilted brane.

 In the state with an arbitrarily bent brane we will
distinguish two components. One  can be viewed as a collection of all
possible Goldstone waves traveling with the speed of light, which
``redshift" away like ordinary matter.
Another component is the vacuum solution (more precisely, a family 
of
solutions) that would be stable if it were not for the  expansion of the
Universe. This configuration ``redshifts" away slower than  matter 
and 
can be
called  the tilted brane configuration (or wrapped, if there are 
several 
windings
on the length of the cylinder $L$). It can only ``redshift" through the 
``stretching"
triggered by the Universe expansion.

In the case of the generic massless fields $\phi_A$ with the flat 
$\eta^{AB}$,
the  solution $\phi = \alpha x$ is stable under any localized 
deformations. If
$\phi_A$'s are the Goldstone bosons arising due to the
spontaneous breaking of some compact symmetry, $\phi_A$'s are 
periodic
(in fact, they are phases defined modulo shifts). This is exactly what
happens in our
case since  the extra dimensions are assumed to be compact. 
Then the solution $\phi = \alpha x$  must be modified
appropriately, see Sec. 2. Here we will add a comment regarding 
 the issue of 
stability. 

To explain the  point we will  use  a simple
example \footnote{The analogy is much more precise than one might
naively  think, since the translation in the external space
can be regarded as an internal U(1) rotation. In the presence of
gravity, this is a gauge rotation gauged by the external components 
of 
the
graviton (graviphoton(s)). We can neglect the coupling to gravity
for the time being, due to the large sizes of extra dimension(s).}
of the  Goldstone field produced as a result of breaking
of some global
$U(1)$ symmetry (a similar model was treated in Sec. 6 
of~\cite{DS1}).
 Start from
the Lagrangian
\begin{equation}
{\cal L} = |\partial_{\mu}\Phi|^2 - {\lambda^2 \over 2}\left(|\Phi|^2 -
v^2\right)^2\, .
\label{othergold}
\end{equation}
The equations of motion have a solution
\begin{equation}
\Phi = c\,e^{i\mu x}\, , \qquad c = \sqrt{v^2 - {\mu^2 \over 
\lambda^2}}\, ,
\label{solution}
\end{equation}
 which corresponds to 
winding of the phase with the period $2\pi / \mu$ as one 
moves along
$x$. This solution has both the  gradient and potential energies (cf. 
the
solutions discussed in~\cite{DS2}). The potential term 
scales as
$\sim \mu^4 / \lambda^2$. It is seen that at $\lambda \rightarrow 
\infty$
the solution (\ref{solution})  becomes pure gradient energy
($v$ must scale accordingly, of course, i.e. $v\sim\lambda^{-1}$). This 
is
because in this limit, the
$x$-dependence can not affect the order parameter and, thus, the 
potential
energy. In the case of  the tilted brane this would mean that the  
brane 
with
a constant tilt carries purely gradient energy in the limit when 
bending
can not affect its
tension. This is true for any brane in the limit when one can ignore 
its
thickness. The corresponding configuration can not decay into the
$\phi$ waves. Its energy
is reduced only through the Friedmann expansion of the Universe 
and, 
thus,
scales
as $\sim a^{-2}$, the scale factor in the Friedmann Universe. This 
means, 
in
turn, that the state at hand will sooner or later dominate over both, 
the
radiation and the matter densities. We will discuss the observational
implications of this fact in Section 5. Before, however, let us discuss
the effect of higher dimensional bulk gravity.

\section{Switching on gravity in the bulk}

The
effective Lagrangian for the zero modes  localized on the brane
becomes
\begin{equation}
{\cal L}_{\rm brane} =
M_{{\rm P}_{\rm f}}^2\, g^{\mu\nu}\left(\partial_{\mu}
\phi_A\partial_{\nu}\phi_B \right) \eta^{AB} +\mbox{fermions}\, ,
\label{Goldstoneslag}
\end{equation}
where $\eta_{AB}$ and $g$ are the external and induced metrics,
respectively. 
The fermion terms are relevant in the supersymmetric case.

So far, we have discussed the brane dynamics on the cylinder 
neglecting
gravity. Now we want to take it into account. In the scenario under
consideration gravity is not confined to the wall, but, rather, 
propagates 
in the
bulk. The components of the graviton belonging to ${\cal M}_4$ and 
${\cal T}^N$
require separate treatment. The effect which is most important
for us is due to  the higher-dimensional components of the
graviton (the so-called graviphotons). For simplicity we will consider 
only a single 
extra
dimension compactified on a circle parametrized by the coordinate 
$X$.
The zero mode component of  $g_{\mu5}$ is the graviphoton $
A_{\mu} (x)$. Viewed as a four-dimensional gauge
field, $A_{\mu} (x)$ gauges the translation in $X$, i.e. $X \rightarrow
X + f(x_{\mu})$, which, from the four-dimensional standpoint, is
an internal U(1) gauge symmetry. Since the brane spontaneously 
breaks 
the
translational invariance, the corresponding gauge symmetry is
realized  nonlinearly. The Goldstone mode $\phi$ is eaten up by the 
gauge
field $A_{\mu}$ (which gets a mass of order  (1
mm)$^{-1}$~\cite{higgseffect}).
As a result, the tilted wall solution $\phi =\alpha x$ we have 
considered
previously  is  pure gauge. It can be compensated by the gauge field 
$A_{x} =
\alpha$ and presents no physically observable effect. Thus, if we 
have 
only one
zero mode on the wall, the tilted wall is indistinguishable from the 
straight one
in the presence of gravity. To make the idea work we must have
two or more zero boson  modes on the wall. Then the tilted wall
(the constant gradient energy solution) will lead to a physically 
observable
excess in $\rho$.

Let us explain this in more detail.
Our solution can be made physical by ``projecting out" the 
graviphoton,
provided the compact manifold breaks translational
invariance in the extra space. The simplest dynamical realization is
as follows.
Consider topologically stable winding configurations~\cite{DS1}.
(Such configurations may anyway be  needed for supersymmetry 
breaking
\cite{DS1} or for stabilizing radii at large distances~\cite{radius}).
Consider a five-dimensional scalar field $\xi(x_{\mu}, X)$
transforming under an internal U(1)$_I$ symmetry as $\xi 
\rightarrow {
e}^{i\alpha} \xi$. Assume that a potential forces the condensate 
$\langle
\xi \rangle \neq 0$ to develop.
The simplest choice can be $V = (|\xi|^2 - v)^2/m$, or any other
similar function. Then the vacuum manifold is a circle, and there are
topologically stable winding configurations
\begin{equation}
  \xi = w\,e^{in X/R}\, , \qquad w = \sqrt{v - n^2m/R^2}  \, ,
\end{equation}
with integer $n$. They  correspond to giving a vacuum expectation 
value  
to
different Kaluza-Klein modes and, therefore, are {\it topologically}
 stable due
to the mapping of the vacuum circle on the external compact space.
Thus we are free to choose any of these states as the ground state.
\footnote{Strictly speaking, we have to restrict ourselves to
$n^2 < vR^2 / m$ since larger winding numbers force the field to
vanish, and the solution will ``unwind". For our purposes, however, it 
is 
sufficient to consider small  enough $n$.}
Configurations with nonzero $n$ break
spontaneously the $X$ translations and, thus, give mass to the 
graviphoton
$A_{\mu}$. As a result, the  graviphoton field can be integrated out 
in 
the 
low-energy effective theory. Then the brane Goldstone $\phi$ 
remains a
physical field. 

More precisely,
the picture is as follows. The theory at hand has two U(1) 
symmetries 
from the  very
beginning:  the internal U(1)$_I$ and ``external"
U(1)$_E$ gauge symmetry under translations in the extra dimension. 
However,
from the
point of view of a four-dimensional observer living on the brane, 
both
of them are  internal symmetries, one global and another gauged by
$A_{\mu}$. The condensate  $\langle
\xi \rangle \neq 0$, with $n \neq 0$,  breaks both U(1)'s 
down to a global U(1) describing the
change of a relative position of the brane in the extra space. The 
brane 
breaks
the latter down to nothing, but since there is no gauge degree of 
freedom left,
the corresponding Goldstone is physical.

A related mechanism of ``projecting out" the graviphoton while still 
keeping at
least one Goldstone mode may be provided by multicomponent walls 
suggested
in Ref.~\cite{SSV}. It was noted that in supersymmetric theories with 
multiple
degenerate vacua, under certain conditions there exists a variety of 
the
BPS-saturated walls. Some of them may coexist together; then, the 
tension of
such ``multilayer" configuration is exactly equal to the sum of the 
tensions of
the individual walls, independently of the distance between the 
layers
(individual walls). A similar effect takes place for the $D$-branes
which are the BPS-states in the limit of unbroken supersymmetry, 
and 
the net force
between two straight parallel branes vanishes (see~\cite{D-branes} 
for a 
review).
This means that, apart from the zero mode corresponding to the 
overall
translation, there are extra zero modes corresponding to shifting the 
layers with
respect to each other, without changing the position of the center of 
mass.
As was mentioned, on the cylinder the exact BPS saturation can be 
achieved
only in the limit $\delta / R\to 0$. Hence, only the first zero mode --
the one related to the overall translations -- is exactly zero, others 
become
quasi-zero. The zero mode is eaten up by the graviphoton, 
eliminating it 
from the
game. The quasi-zero modes remain physical, their dynamics is 
described by
a low-energy Lagrangian. The supersymmetry breaking in general 
will 
give
mass to this pseudo-Goldstones since the net force between branes 
(or
walls) is non-zero. 

If there are no other massless states in the bulk,
at  large distances the only force between the branes is due to 
gravity 
which gives a small $r^N$-suppressed mass to the Goldstones ($r$ is 
the
interbrane separation). This mass can be made arbitrarily 
small if the separation is
large and, practically, it can be neglected. All dynamical 
consequences 
are
the same as discussed in Sec. 3.

\section{Implications for cosmology}

 Now let us discuss  cosmology
in more detail. We will treat the problem  from the standpoint  of
the effective four-dimensional field theory at scales $\gg\gg R$.
As was said above, in this picture the fluctuations of the brane in the
external space $\phi_A(x_{\mu})$ can be described by the Goldstone 
modes 
which, 
at energies $\ll T$ (almost flat brane), behave as free particles.
Canonically normalized fields are $\chi_A \sim (T)^{1/4} \phi_A$.

For simplicity, consider a fluctuation in one transverse direction only.
Then, very roughly, the issue is reduced to the behavior
of a free field $\chi$ in an expanding Universe. The question is what 
are
the initial conditions for such a field? The general solution of its
equation of motion (the comoving coordinates are assumed)
\begin{equation}
\partial^2 \chi = 0
\label{freeequation}
\end{equation}
has the form 
\begin{equation}
\chi = \sqrt{T} c x  + \chi_p { e}^{ipx}  
\label{waves}
\end{equation}
(modulo the Lorentz transformations), 
where $c$ is a constant and the second term is some collection
of massless plane waves. The energy stored in the second term will 
just
``redshift" away, like massless matter.
However, the
first term produces a rather different contribution. In the present 
context
this term  describes the tilted brane with the  tilt angle given by $c 
\sim
R/r$. In other words,  when we move along $x$, the brane is 
wrapping 
around
the extra dimension with the period $\sim r$. In the absence of 
gravity, 
in the
limit of an infinitely thin brane, 
this is a stable configuration for any $c$. In the presence of gravity,
however,  its energy ``redshifts" away as $\sim a^{-2}$,  due to the 
Friedmann
expansion. Therefore,  eventually this energy will dominate over  
matter. 
When this  happens actually depends on the initial condition for
$r (c)$ and on the subsequent evolution of the scale factor.
Assume
that initially  $c \sim RH_{\rm in}$, where $H_{\rm in}$ is the 
Hubble 
parameter
at
that time. In  other words, we assume that, when the brane was 
``formed",
it wrapped around the extra dimension once ( on average)  per the 
causally
connected region $\sim H_{\rm in}^{-1}$. Now, at present time, this 
region
must have had evolved into a volume comparable to the present 
Hubble 
size (or
larger). This is required by  whatever mechanism  solving the 
horizon
problem. This means that
\begin{equation}
r_{\rm today} \sim \frac{a_{\rm today}}{H_{\rm in}a_{\rm in} 
}\gsim
H^{-1}_{\rm today}\, .
\end{equation}
Thus, the energy density of a tilted brane would be $ \lsim \rho_c$.  
In particular, it is sufficient to have 
a period of inflation
with the  number of $e$-foldings
\begin{equation}
N \simeq {\rm ln}{H_{\rm in}M_P \over T_{\rm today} T_R}
\, ,
 \label{ef}
\end{equation}
and the subsequent reheating temperature $T_R$, where $T_{\rm 
today} \sim 3
K^o$. For instance, a brief period of the ``brane
inflation" \cite{braneinflation} can
do the job.\footnote{The wrapped brane can provide an additional
(time-dependent) force stabilizing $R$. This may have implications
for the early cosmology  \cite{radiuscosmology}.}

In reality, however, we expect inflation to have more $e$-foldings 
and,
thus, $\rho_{\rm brane} \lsim \rho_c$ is rather natural.

\section{The Lorentz Symmetry Breaking}

The important fact is that the solution at hand spontaneously breaks 
the 
Lorentz
and rotational invariances in four dimensions. 
This would result in a global anisotropy in the expansion if 
$\rho_{\rm
brane}$ were to dominate
the Universe. 
Thus, we must require $\rho_{\rm brane}$ to be subdominant
today, but it can become dominant at any time in the future. In this 
respect any
observational evidence of a global anisotropy would be extremely 
important for constraining $\rho_{\rm brane}$.

 The tilted brane would induce
rotational (or Lorentz) noninvariant terms in the effective
four-dimensional standard-model Lagrangian. The important fact is 
that 
the
brane Goldstones (or pseudo-Goldstones,  which are similar in this 
respect)
necessarily couple to all  particles living on the brane through an
induced metric on the brane. Thus, the operators of the form
\beq
 \frac{\partial \phi}{\partial x_\mu} \,
\frac{\partial \phi}{\partial x^\nu}\, \psi\gamma_{\mu}\partial 
x^\nu\psi
+ \frac{\partial \phi}{\partial x_\mu} \,
\frac{\partial \phi}{\partial x^\nu} \, F_{\mu\alpha}F^{\nu\alpha}
+...
\eeq
will appear (accompanied by the appropriate powers of $T$). Here, 
$\psi$ are the  matter fermions and $F_{\mu\alpha}$ 
stands
for the matter gauge fields. 
 In the background of a titled brane $\phi = \alpha x$, these terms 
will
induce effective Lorentz-violating interactions among the 
standard-model
fields.

Can one have tilted branes and still avoid the anisotropy and 
violations of the rotational invariance?
In principle, the answer is positive. To this end one must deal
with more complicated manifolds, such that  the topology of the 
manifold ${\cal
M}^\prime_N$ is the same as that of our $3$-dimensional space 
${\cal 
M}_3$
(of course, in any case, locally ${\cal M}_3$ must be Minkowskian).
In other words,  ${\cal M}^\prime_N$ must include unshrinkable 
surfaces $C_3$  that
are isomorphic to ${\cal M}_3$. In this  case the brane that 
corresponds 
to  a
point
on $C_3$ can move around when one moves along the brane in ${\cal 
M}_3$.
The topologically nontrivial configuration of interest emerges when 
this
motion maps $C_3$ onto ${\cal M}_3$. For such a configuration, one 
can 
find a preferred reference frame  for which rotations on ${\cal M}_3$ 
will not be 
broken,
but the Lorentz invariance will be broken in the arbitrary reference 
frame.  To be more specific, let us give an example.
Imagine that we have an external space, with three extra 
dimensions, which
forms a manifold $K$ of radius $R$ (say, one can think of  $K$
as of  a 
three-sphere). Assume that ``our" 
three-dimensional
space is also a very  large
manifold $K$, its radius is much larger than $H^{-1}$.
Let us call these two manifolds ``small" and 
``big", respectively. Clearly,  at human scales the ``large" manifold
(plus time)  is
identical to ${\cal M}_4$, but not globally.
Now, our brane corresponds to a point on the  ``small" $K$. Imagine a
 configuration such that when one moves  in ``large" $K$, the 
point
in ``small" $K$ also moves in the same way. Thus, we get a
mapping $K \to K$,
which is certainly topologically stable. Roughly speaking, the brane 
wraps
isotropically in all
directions. Such a wrapping is isotropic and stable.

What is the low-energy picture corresponding to this construction?
Since $K$ has tree dimensions, the brane breaks three translational 
invariances. 
Thus,
there are three massless Goldstone modes in the effective 
low-energy
theory, $\Phi^A\, , \,\,\,
A=1,2,3$. Let $x^A$ be three coordinates on our large $ S_3$ which
locally look as the Cartesian coordinates in  our Minkowski space.
Then, the solution is $\Phi^A = \alpha x^A$.
 This solution is isotropic because of the spherical symmetry of the
problem. Its energy density will still scale  $\propto 1/a^2$
because this  is essentially the same  gradient energy solution we 
have
discussed previously. 

\section{Conclusions}

The idea of confining our Universe to a wall which ensures an 
appropriate
supersymmetry breaking~\cite{DS1} seems to be promising. At the 
very 
least, it
deserves further investigation.  Being combined with the idea of 
compactification of
the extra dimensions and allowing gravity to propagate in the 
bulk~\cite{add}
it leads to potentially realistic and reach phenomenology.  In this
paper we have
shown that the walls on  ${\cal M}_4\times {\cal M}^\prime_N$ 
where 
${\cal
M}^\prime_N$ is compact
generate peculiar theoretical effects due to tilting. 
The situation becomes especially interesting when gravity is 
switched 
on.
The appropriate theoretical framework for its analysis is provided by 
the
effective low-energy theories of the Goldstone modes on the
brane. After  gravity is switched on one of these modes is eaten up 
by 
the
graviphoton (making it massive and eliminating it from the massless 
particle
spectrum).  We presented models where there are residual physical
Goldstone (or pseudo-Goldstone) modes. 

These solutions produce a framework for the spontaneous breaking 
of 
the Lorentz
and rotational invariance and may have observable consequences.

\vspace{0.5cm}

{\bf Acknowledgments}: \hspace{0.2cm} 
We would like to thank A. Dolgov,  I. Kogan, K. Olive,  T. Piran, and A. 
Vilenkin for useful discussions and comments.
This work was supported in part by DOE under the grant number
DE-FG02-94ER40823.


\begin{thebibliography}{99}

\bibitem{DS1}
G. Dvali and M. Shifman, {\it Nucl. Phys.} {\bf B504} (1997) 127.

\bibitem{DS2} 
G. Dvali and M. Shifman, hep-th/9901111 (to appear in Phys. Lett. B).

\bibitem{add} N. Arkani-Hamed, S. Dimopoulos, and  G. Dvali, {\it 
Phys.
Lett.} {\bf B516}  (1998) 70.

\bibitem{AADD} I. Antoniadis, N. Arkani-Hamed, S. Dimopoulos, and 
G.  Dvali,  {\it Phys. Lett.} {\bf B436} (1998) 257.

\bibitem{lw} E. Witten, {\it Nucl. Phys.} {\bf B471} (1996) 135;
J. Lykken, {\it Phys. Rev.} {\bf D54} (1996) 3693.

\bibitem{ddg}  Lowering the GUT scale was suggested in, K.R. Dienes, 
E.
Dudas and T. Gherghetta, Phys. Lett. {\bf B436} (1998) 55;
hep-ph/9806292.

\bibitem{localization1/2} V.A. Rubakov and M.E. Shaposhnikov, {\it 
Phys. Lett.}
{\bf B125} (1983) 136;

\bibitem{localization1} 
G. Dvali and M. Shifman, {\it Phys. Lett.}
{\bf B396} (1997) 64; Erratum {\it Phys. Lett.}
{\bf B407} (1997) 452.  

\bibitem{D-branes} J. Polchinski, {\em String Theory}, (Cambridge 
University Press, 1998).

\bibitem{ST} G. Shiu and S.-H.H. Tye, {\it Phys. Rev.} {\bf D58} (1998) 
106007,
hep-th/9805147;\\
Z. Kakushadze and S.-H.H. Tye, hep-th/9809147.

\bibitem{ignatios} I. Antoniadis and B. Pioline, hep-th/9902055;

\bibitem{braneinflation} G. Dvali and H. Tye,  hep-ph/9812483.

\bibitem{gigad} G. Dvali and G. Gabadadze, hep-ph/9904221.

\bibitem{ovrut} A. Lukas, B. A. Ovrut, and D. Waldram, 
hep-th/9806022;
hep-th/9902071.

\bibitem{HW} P. Ho\v{r}ava  and E. Witten, {\it Nucl. Phys.}
{\bf B475} (1996) 94. 
 
\bibitem{mmtr} M. Maggiore and  A. Riotto, hep-th/9811089.

\bibitem{ZKO}
Ya.B. Zeldovich, I.Yu. Kobzarev, and L.B. Okun,
{\it Zh. Eksp. Teor. Fiz.} {\bf 67} (1974) 3 [{\it Sov. Phys. JETP} {\bf 
40} (1974) 1].

\bibitem{higgseffect} N. Arkani-Hamed, S. Dimopoulos, and G. Dvali, 
hep-ph/9803315.

\bibitem{radius} R. Sundrum, hep-ph/9807348; 
N. Arkani-Hamed, S. Dimopoulos, and J. March-Russell, 
hep-th/9809124.

\bibitem{SSV}
M. Shifman, {\it Phys. Rev.} {\bf D57} (1998) 1258;
 M. Shifman and  M. Voloshin, {\it Phys. Rev. } {\bf D57} (1998) 
2590.

\bibitem{radiuscosmology} N. Kaloper and A. Linde, 
hep-th/9811141;
A. Mazumdar, hep-ph/9902381;  N. Arkani-Hamed, S. Dimopoulos,
N. Kaloper, and J. March-Russell, hep-hp/9903239; hep-ph/9903224.

\end{thebibliography}
\end{document}